\documentclass[aps,prd,preprint,superscriptaddress,tightenlines,nofootinbib]{revtex4}

\usepackage{graphicx}% Include figure files
\usepackage{dcolumn}% Align table columns on decimal point
\usepackage{bm}% bold math
\usepackage{epsfig}

\begin{document}

%===================> ADD here your LATEX definitions

%\newcommand{\sprod}{$S_{1} \times S_{2}$ }
%\newcommand{\zz}{Z$^{0}$}
\newcommand{\ee}{e$^+$e$^-$}
\newcommand{\ff}{f$_{2}$(1525)}
\newcommand{\bb}{$b \overline{b}$}
\newcommand{\cc}{$c \overline{c}$}
\newcommand{\sbs}{$s \overline{s}$}
\newcommand{\uu}{$u \overline{u}$}
\newcommand{\dd}{$d \overline{d}$}
\newcommand{\qq}{$q \overline{q}$}
\newcommand{\suo}{\rm{\mbox{$\epsilon_{b}$}}}
\newcommand{\loro}{\rm{\mbox{$\epsilon_{c}$}}}
\newcommand{\kos}{\ifmmode \mathrm{K^{0}_{S}} \else K$^{0}_{\mathrm S} $ \fi}
\newcommand{\kol}{\ifmmode \mathrm{K^{0}_{L}} \else K$^{0}_{\mathrm L} $ \fi}
\newcommand{\ko}{\ifmmode {\mathrm K^{0}} \else K$^{0} $ \fi}

%\newcommand
%{\kob}
%{\ifmmode {\overline{\mathrm{K}^{0}}} \else $\overline{\mathrm{K}^{0}}$\fi}
\def\tpc{three-particle correlation}
\def\twopc{two-particle correlation}
\def\ksks{K$^0_S$K$^0_S$}
\def\ee{e$^+$e$^-$}
\def\ff{f$_{2}$(1525)}

%\preprint line(s) will be ignored for PRL/PRD

\hfill {CLNS 03/1853} 

\hfill {CLEO 03-14} 

%\hfill {To be submitted to Phys. Rev. Lett.} 

%\preprint{CLNS-03-1853} % For paper draft CBX YY-NN -> Draft YY-NNA
%\preprint{CLEO 03-14}         % for CLNS notes
%\preprint{To be submitted to Phys. Rev. Lett.}      % For conference papers

\title{Observation of $\eta_{c}^{\prime}$ Production in $\gamma\gamma$ Fusion
at CLEO~}
% for conference papers (ask CLEOAC for appropriate text)
%\thanks{Submitted to the
%International Europhysics Conference on High Energy Physics,
%July 2003, Aachen}

%-------- INSERT HERE ------------
% Your author list goes here  REMOVE EVERYTHING to END INSERT and
% replace with your authorlist (ask cleoac).
%
\author{D.~M.~Asner}
\author{S.~A.~Dytman}
\author{S.~Mehrabyan}
\author{J.~A.~Mueller}
\author{S.~Nam}
\author{V.~Savinov}
\affiliation{University of Pittsburgh, Pittsburgh, Pennsylvania 15260}
\author{G.~S.~Huang}
\author{D.~H.~Miller}
\author{V.~Pavlunin}
\author{B.~Sanghi}
\author{E.~I.~Shibata}
\author{I.~P.~J.~Shipsey}
\affiliation{Purdue University, West Lafayette, Indiana 47907}
% ...
%\author{(CLEO Collaboration)} %FOR PRD_SPECIAL_CHANGEME
\collaboration{CLEO Collaboration} %FOR PRL,CLNS
\noaffiliation
%-------- END INSERT ------------

%please hard code the date when you have a final draft and submit to CLEOAC

\date{December 23, 2003}

\begin{abstract}

We report on the observation of the
$\eta_{c}^{\prime}$(2$^{1}S_{0}$),
the radial excitation of $\eta_{c}$(1$^{1}S_{0}$)
ground state of charmonium, in the two-photon fusion reaction
$\gamma\gamma \rightarrow \eta_{c}^{\prime}
\rightarrow K_{S}^{0}K^{\pm}\pi^{\mp}$
in 13.6 fb$^{-1}$ of CLEO II/II.V data
and 13.1 fb$^{-1}$ of CLEO III data. We obtain 
$M(\eta_{c}^{\prime})$=3642.9$\pm$3.1(stat)$\pm$1.5(syst) MeV, and
$M(\eta_{c})$=2981.8$\pm$1.3(stat)$\pm$1.5(syst) MeV.
The corresponding values of hyperfine splittings between
$^{1}S_{0}$ and $^{3}S_{1}$ states are
$\Delta M_{hf}(1S)$=115.1$\pm$2.0 MeV, $\Delta M_{hf}(2S)$=43.1$\pm$3.4 
MeV. Assuming that the $\eta_{c}$ and $\eta_{c}^{\prime}$ have equal 
branching fractions to $K_{S}K\pi$, we obtain 
$\Gamma_{\gamma\gamma}(\eta_{c}^{\prime})$=1.3$\pm$0.6 keV.

\end{abstract}


\maketitle

Quantum Chromodynamics (QCD) is the accepted theory of the strong
interaction. Charmonium ($c\bar{c}$) states provide an excellent laboratory 
for the study of the QCD interactions. 
Experimental data are generally compared with perturbative
predictions with the QCD interaction modeled by a potential. 
The central part of the popular Cornell potential~\cite{cornell} 
consists of a one--gluon 
exchange ``Coulombic'' part $\propto$~1/$r$, and a ``confinement''
part $\propto$~$r$. The spin--dependence of this potential,
with spin--orbit, spin--spin and tensor components, is generally assumed to
arise only from the vector Coulombic part. The confinement potential is
assumed to be scalar with only a minimal spin--orbit contribution due to 
Thomas precession. There is little experimental evidence to support the
assumption of the pure Lorentz scalar nature of the confinement potential.
One of the best ways to study the validity of this assumption is
to measure the hyperfine splitting of states which sample the confinement
region of the  $q\bar{q}$ potential. The 2S states of charmonium,
the $\psi^{\prime}$(2$^{3}S_{1}$) and  $\eta_{c}^{\prime}$(2$^{1}S_{0}$),
are ideal for this purpose. The mass of the  $\psi^{\prime}$ is known very 
precisely, $M(\psi^{\prime})$=3685.96$\pm$0.09 MeV~\cite{pdg2002}, 
but the $\eta_{c}^{\prime}$ has not been firmly identified until recently.
In this Letter we report on the observation of the $\eta_{c}^{\prime}$
in independent CLEO II and CLEO III measurements of the two-photon 
fusion reaction
\begin{equation}
e^{+}e^{-} \rightarrow e^{+}e^{-} (\gamma\gamma),
~~~\gamma\gamma  \rightarrow \eta_{c}^{\prime} 
 \rightarrow K_{S}^{0}K^{\pm}\pi^{\mp} \, .
\end{equation}

In 1982 the Crystal Ball collaboration reported the observation
of a small enhancement at  $E_{\gamma}$~$\approx$~91 MeV in the
inclusive photon spectrum from the reaction 
$e^{+}e^{-} \rightarrow \psi^{\prime}  \rightarrow \gamma X$,
and interpreted it as due to  $\eta_{c}^{\prime}$ with
 $M(\eta_{c}^{\prime})$=3594$\pm$5 MeV,  
$\Gamma(\eta_{c}^{\prime})$$<$8 MeV~[1,2]. 
 This observation, which 
corresponds to a 2S hyperfine splitting 
$\Delta M_{hf}$(2S)=$M(\psi^{\prime})-M(\eta_{c}^{\prime})$=92$\pm$5 MeV, was
in qualitative accord with the well established 1S hyperfine splitting,
$\Delta M_{hf}$(1S)=$M(J/\psi)-M(\eta_{c})$=117$\pm$2 MeV~\cite{pdg2002}.
However, it was not confirmed, and the listing
of the  $\eta_{c}^{\prime}$ was dropped by the PDG~\cite{pdg2002}
from the meson summary list. 
The Fermilab experiments E760/E835~\cite{e760}
failed to identify  $\eta_{c}^{\prime}$ 
in the reaction 
$\bar{p}p~\rightarrow~\eta_{c}^{\prime}~\rightarrow \gamma\gamma$,
for  $\eta_{c}^{\prime}$ 
mass in the range $M(\eta_{c}^{\prime})$=3575--3660 MeV. 
Similarly, in $e^{+}e^{-}$ collisions at $\sqrt{s}$ $\approx$ 91 GeV
 DELPHI~\cite{delphi}, and later L3~\cite{l3},
found no evidence for $\eta_{c}^{\prime}$ 
in the reaction $\gamma\gamma~\rightarrow~hadrons$,
in the mass range, 3500--3800 MeV, and concluded
that its population in this reaction was less than a third of that
of the $\eta_{c}$. 
A recent preliminary CLEO 
measurement~\cite{cleo_psip} of the inclusive photon spectrum 
from $\psi^{\prime}  \rightarrow \gamma X$ has also not found any evidence
for the excitation of  $\eta_{c}^{\prime}$.

The theoretical situation was equally uncertain. The perturbative
prediction for the hyperfine splitting of the S states of charmonia is,
in the lowest order
\begin{equation}
\Delta M_{hf}(S)=[32\pi\alpha_{s}/(9m_{c}^{2})]|\Psi(0)|^{2}  \,.
\end{equation}
Thus, assuming that the strong coupling constant 
$\alpha_{s}(2S)=\alpha_{s}(1S)$,
\begin{displaymath}
%\begin{center}
{\Delta M_{hf}(2S) \over \Delta M_{hf}(1S)}=
{|\Psi(0)/m_{c}|_{2S}^{2} \over |\Psi(0)/m_{c}|_{1S}^{2}}=
{\Gamma(\psi^{\prime} \to e^{+}e^{-}) \over \Gamma(J/\psi \to e^{+}e^{-})}
{M^{2}(\psi^{\prime}) \over M^{2}(J/\psi)} \, 
%\left({M(\psi^{\prime}) \over M(J/\psi)}\right)^{2} \, 
%\end{center}
\end{displaymath}
since $\Gamma(^{3}S_{1} \to e^{+}e^{-})$ is proportional to
$|\Psi(0)|^{2}/M^{2}(^{3}S_{1})$. Substituting experimental 
values~\cite{pdg2002} yields,
$\Delta M_{hf}(2S)$=68$\pm$7 MeV.
Buchm\"uller and Tye~\cite{tye} have pointed out that in order to
take approximate account of binding energy, $m_{c}$ in Eq. (2)
can be replaced by $M(^{3}S_{1})$/2, which leads to $\Delta M_{hf}(2S)$=48$\pm$5 MeV.

Numerous potential model predictions for 
$\Delta M_{hf}(1S,2S)$ exist. Most of them make the assumption that
the confinement potential is scalar. The predictions range from
$\Delta M_{hf}(2S)$=60--100 MeV. A recent calculation with a screened 
Coulombic potential~\cite{riska} predicts 
$\Delta M_{hf}(2S)$=38 MeV,
but it gives  splittings for the $^{3}P_{J}$ states which are
factor of two smaller than experimentally measured.
Two recent quenched lattice calculations predict 
$\Delta M_{hf}(2S)$=94--106 MeV~\cite{bali}, and
$\Delta M_{hf}(2S)$=25--43 MeV~\cite{lattice}, respectively. 
Most predictions make
the caveat that coupled-channel effects, which were not included, 
may be important for $\Delta M_{hf}(2S)$ because of the proximity
of the 2S states to the $D\bar{D}$ threshold at 3.73 GeV.

The first reports of a successful identification of  $\eta_{c}^{\prime}$
came recently from two measurements by the Belle collaboration.
In the decay of 45 million $B$ mesons, $B\rightarrow~K(K_{S}K\pi)$, 
they observed peaks in the $K_{S}K\pi$ invariant mass spectrum 
corresponding to the $\eta_{c}$ and $\eta_{c}^{\prime}$, and reported   
$M(\eta_{c}^{\prime})$=3654$\pm$6$\pm$8 MeV~\cite{belle1}~\cite{errors}. 
They also reported~\cite{belle2} 
$\eta_{c}^{\prime}$ observation in double charmonium
production, $e^{+}e^{-}\rightarrow~J/\psi+\eta_{c}^{\prime}$,
in 46.2 fb$^{-1}$ of $e^{+}e^{-}$ data at $\sqrt{s}$~$\approx$~10 GeV. 
They reported $M(\eta_{c}^{\prime})$=3622$\pm$12(stat) MeV.
The fact that both masses were significantly larger than that
reported by the Crystal Ball collaboration provided for great 
interest in 
confirming the $\eta_{c}^{\prime}$ observation in independent measurements
at CLEO. 

At CLEO we had earlier reported~\cite{bran} the identification and study of 
$\eta_{c}(^{1}S_{0})$ in the two-photon fusion reaction of Eq. (1)
in 13.6 fb$^{-1}$ of CLEO II data at the $\Upsilon$(4S) and vicinity.
We have  reanalyzed CLEO II data 
with the resonance search extended for $M(K_{S}K\pi)$ up to 4.1 GeV. A
positive signal was observed for an $\eta_{c}^{\prime}$ mass of 
$\sim$3643 MeV. 
In order to confirm this observation,  13.1  fb$^{-1}$ of data taken
at, and in the vicinity of, the $\Upsilon$(1S$\to$4S) resonances 
with the improved CLEO III detector were analyzed. Results which were
consistent with those from the CLEO II data were obtained. 

Charged particle tracking and dE/dx measurements in the 
CLEO II~\cite{kubota,hill} detectors were done by 
various concentric devices (straw tube chamber, drift chamber,
and silicon vertex detector) operating in a 1.5 T superconducting solenoid.
They  have been described in detail in Ref.~\cite{bran}. 
$K_{S}^{0}$ were uniquely reconstructed
from the displaced vertex of their $\pi^{+}\pi^{-}$ decay~\cite{bran}.
$K$/$\pi$ separation was done by using the dE/dx and TOF 
information.

For CLEO III~\cite{cleo3}, the charged particle tracking system
was replaced with four layers of double-sided silicon detectors,
surrounded by a new, 47-layer drift chamber~\cite{dr3}.
The CLEO II time-of-flight system was replaced by a
ring-imaging Cherenkov detector (RICH)~\cite{rich} 
which distinguishes $K^{\pm}$ from $\pi^{\pm}$ over 80\% of 
solid angle. The two charged tracks not from the $K_{S}^{0}$ decay 
were tested as being
either kaons or pions. All events were used in which the charged particle
candidate is identified as a $K$ or $\pi$ by the RICH detector. For 
$p$~$>$~2 GeV/c (as for most of our $K^{\pm}$ candidates)
the RICH identifies kaons with efficiency greater than 81\% while having
less than 2\% probability of a pion faking a kaon.
When  $K$/$\pi$ discrimination by RICH was not possible,
dE/dx measurements from the drift chamber were used.  
For $1 < p < 2$ GeV/c, however,  $K$/$\pi$ separation was 
difficult using dE/dx, and such events were rejected.

In order to insure production via two-photon fusion and only four 
charged particles in the event, additional cuts were made in total  
transverse momentum $P_{T}$ of the $K_{S}K\pi$ system, and in 
neutral energy  E$_{neut}$ not associated with the charged particles.

\begin{figure}[hbt]
\vspace*{-0.3cm}
\label{fg:fits}
\begin{center}
\includegraphics*[width=3.2in]{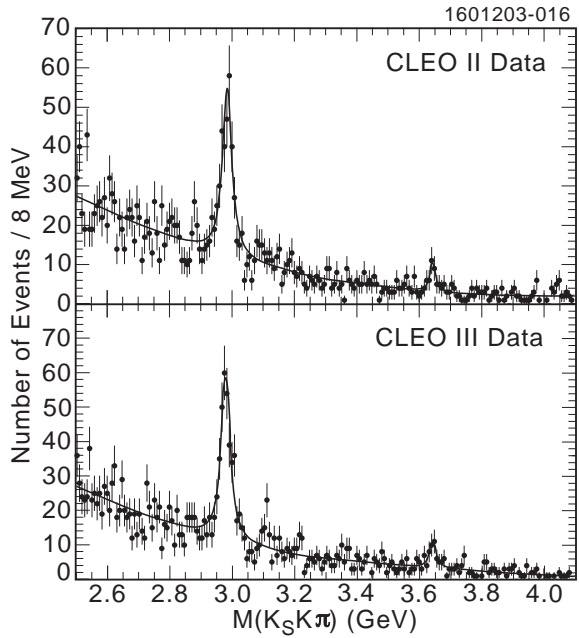}
\end{center}
\vspace*{-0.6cm}
\caption{Invariant mass distributions for $K_{S}^{0}K^{\pm}\pi^{\mp}$ events 
from (top) the CLEO II data and (bottom) the CLEO III data. The curves in the
figures are results of fits described in the text.}
\end{figure}

The Monte Carlo simulation of the CLEO detector
response was based upon GEANT~\cite{geant}, with events for
reaction (1) simulated using the formalism of 
Budnev {\it et al.}~\cite{budnev}. Simulated events were processed in
the same manner as the data to determine the $K_{S}^{0}K^{\pm}\pi^{\mp}$
detection efficiency. Efficiencies for trigger,  $K_{S}^{0}$ identification,
four--charged track reconstruction,  $\pi$/$K$ identification, and cuts
on  $P_{T}$ and  E$_{neut}$ were determined. The overall detection
efficiencies ($\epsilon$) obtained from large
statistics Monte Carlo samples are listed in Table I.

The  $K_{S}^{0}K^{\pm}\pi^{\mp}$ invariant mass
plots using the event selections for CLEO II and CLEO III
are shown in Fig.~1.
Clear enhancements at masses $\sim$2982 and $\sim$3643 MeV
are visible in both, which we label as 
$\eta_{c}$ and $\eta_{c}^{\prime}$, respectively.
There is a some evidence for a small enhancement
at $\sim$3.1 GeV, presumably from the population of $J/\psi$
via initial state radiation, but it is found that its inclusion
has no effect on our best fit parameters.

In order to extract numerical results from these data we
have made maximum likelihood fits to these
spectra using a polynomial background and two
resonances with Breit--Wigner parametrization, convoluted with
double Gaussians representing the experimental resolution functions
as obtained from Monte Carlo simulations. 
For CLEO II data the widths (relative magnitudes) of the Gaussians
were $\sigma_{1}$=10.0 MeV (78\%),  $\sigma_{2}$=37.2 MeV (22\%).
For CLEO III data the corresponding numbers were
 $\sigma_{1}$=8.7 MeV (74\%),  $\sigma_{2}$=26.6 MeV (26\%).

As described in our previous publication~\cite{bran}, most of the observed  
background under our signal events is expected to arise from non
$\gamma\gamma \rightarrow K_{S}^{0}K^{\pm}\pi^{\mp}$ 
sources such as events with at least one missing $\pi^{0}$ or
$\gamma$,
as well as events of the type  $e^{+}e^{-}~\rightarrow$~hadrons,
$e^{+}e^{-}~\rightarrow~\tau^{+}\tau^{-}$,
and $\gamma\gamma~\rightarrow~\tau^{+}\tau^{-}$. We have therefore
not taken into account possible interference between resonance and 
continuum in fitting our data. The results presented
in Table I are based on our final event selections, $P_{T}$$<$0.6 GeV, 
E$_{neut}$$<$0.2 GeV (CLEO II), or E$_{neut}$$<$0.4 GeV (CLEO III),
second-order polynomial backgrounds, and separate fits in the mass regions 
2.5--3.5 GeV($\eta_{c}$), and 3.3--4.1 GeV($\eta_{c}^{\prime}$).
The ``significance levels'' of the enhancements listed in Table I
were obtained as
$\sigma\equiv$$\sqrt{-2ln(L_{0}/L_{max})}$,
 where $L_{max}$ is the maximum likelihood value from the fits
described above, and  $L_{0}$ is the likelihood value from the fits
with either no $\eta_{c}$ or no $\eta_{c}^{\prime}$ resonance. Combining
the independent significance levels of the CLEO II and CLEO III measurements 
in quadrature gives the significance level for our observation of 
$\eta_{c}^{\prime}$ as 6.5$\sigma$.

\begin{table}[hbt]
\caption{Summary of the results  for $\eta_{c}$ and $\eta_{c}^{\prime}$
for both CLEO II and CLEO III data sets. The errors shown are
statistical only.}
\label{tb:results}
\begin{center}
\begin{tabular}{|c|c|c||c|c|} \hline
& \multicolumn{2}{c||}{CLEO II} &  \multicolumn{2}{c|}{CLEO III} \\
\hline
   &  $\eta_{c}$ &  $\eta_{c}^{\prime}$ &
$\eta_{c}$ &  $\eta_{c}^{\prime}$ \\ \hline
$\epsilon$ (\%) & 10.0 & 13.8 & 8.9 & 11.9 \\
$N$, events & 282$\pm$30 &28$^{+13}_{-10}$ & 310 $\pm$29&
33$^{+14}_{-11}$\\
$M$(MeV) &  2984.2$\pm$2.0 &3642.4$\pm$4.4 & 2980.0$\pm$1.7&
3643.4$\pm$4.3\\
$\Gamma$(MeV) &  24.7$\pm$5.1 &3.9$\pm$18.0 & 24.8$\pm$4.5 & 8.4 $\pm$17.1
\\
signif.($\sigma$) & 15.1 & 4.4 & 17.0 & 4.8 \\
\hline \hline
$R(\eta_{c}^{\prime}/\eta_{c})$ & \multicolumn{2}{c||} 
{0.17$\pm$0.07}  & \multicolumn{2}{c|}{0.19$\pm$0.08} \\ \hline 
\end{tabular}
\end{center}
\end{table}

Photon-photon fusion is expected to populate positive charge
conjugation resonances mainly when the photons are almost real,
{\it i.e.,} when the transverse momenta of both of them, and therefore
of the sum of final state particles is small.
In order to test whether the
observed $\eta_{c}^{\prime}$ peaks are
due primarily to two-photon events,
we have examined the production of $\eta_{c}^{\prime}$ in several
subregions of transverse momentum. We find that both
the CLEO II and CLEO III $P_{T}$ distributions  
are statistically consistent~\cite{eps} with the expectations from our
two-photon Monte Carlo simulations~\cite{geant,budnev}, and we 
conclude that in both data sets the $\eta_{c}^{\prime}$ peak is mainly due 
to two-photon fusion.

It is of interest to compare the two-photon partial width of 
$\eta_{c}^{\prime}$ to that of $\eta_{c}$. The quantity 
that can be directly obtained from the data is

\begin{displaymath}
%\begin{center}
R(\eta_{c}^{\prime}/\eta_{c})\equiv{\Gamma_{\gamma \gamma}(\eta_{c}^{\prime}) \times {\cal B}(\eta_{c}^{\prime} \to
K_{S}K\pi) \over \Gamma_{\gamma \gamma}(\eta_{c}) \times
{\cal B}(\eta_{c} \to K_{S}K\pi)} \, .
%\end{center}
\end{displaymath}

In terms of the measured quantities

\begin{displaymath}
%\begin{center}
R(\eta_{c}^{\prime}/\eta_{c})=
{N(\eta_{c}^{\prime}) \over N(\eta_{c})}
\times{\Phi(m_{\eta_{c}}) \over \Phi(m_{\eta_{c}^{\prime}})}
\times{ \epsilon(\eta_{c}) \over \epsilon(\eta_{c}^{\prime})} \,~~~.
%\end{center}
\end{displaymath}

$\Phi(m_{\eta_{c}})/\Phi(m_{\eta_{c}^{\prime}})$=2.40$\pm$0.05 is the ratio of
the two-photon fluxes at the $\eta_{c}$ and  $\eta_{c}^{\prime}$ 
masses~\cite{budnev}.  This leads to results for 
$R(\eta_{c}^{\prime}/\eta_{c})$ given in Table I.

We have attempted to determine the uncertainty in our mass measurements
due to the calibration of our mass scale
by comparing the masses we measure from our data for 
$K_{S}^{0}(\to \pi^{+}\pi^{-})$,  $D^{0}(\to K_{S}^{0}\pi^{\pm}\pi^{\mp})$,
and $D^{\pm}(\to K^{\pm}\pi^{\pm}\pi^{\mp})$ with their
known values~\cite{pdg2002}. We estimate this uncertainty
to be $\le$ 1 MeV in the $\eta_{c}$ and  $\eta_{c}^{\prime}$
mass regions for both CLEO II and CLEO III data.
Systematic uncertainties may also arise due to the fitting procedures
for the invariant mass spectra. We find that the different choices of the
background parametrization (polynomials, power--law, or exponential)
and peak shape parametrization lead to variations in mass of 
$\le$ 0.5 MeV. It is also found that Monte Carlo events have a reconstructed 
invariant $K_{S}^{0}K^{\pm}\pi^{\mp}$ mass that differs from the input mass 
at levels $\le$ 1 MeV.

We consider the above contributions as being independent of each
other, and by combining them in quadrature, we obtain
a conservative estimate of possible systematic bias in the
$\eta_{c}$ and $\eta_{c}^{\prime}$ masses to be 1.5 MeV for 
both CLEO II and CLEO III.

        Using high statistics samples of $D$ mesons and the larger
$\eta_c$ samples, we have checked that variations in particle identification 
and event selection criteria do not give rise to changes in our results 
in a statistically significant way. 

The dominant source of systematic uncertainty in the determination
of total widths, two-photon widths and the ratio $R$ is found
to be the choice of the background shape. 

The present analysis of the CLEO II data (Table I), including the
systematic errors, yields
$M(\eta_{c})$=2984.2$\pm$2.0$\pm$1.5 MeV. In our earlier
publication for the same data we reported~\cite{bran} 
$M(\eta_{c})$=2980.4$\pm$2.3$\pm$0.6 MeV. A careful examination
of the event selection used there has revealed that an 
algorithm used for charged track identification
led to the inclusion of some ($\sim$13\%) false and poorly
measured events.
Rejection of these events is the main reason for the larger 
mass obtained here. The present determination 
supersedes the earlier reported mass value. The present
analysis of CLEO II data also yields 
$\Gamma(\eta_{c})$=24.7$\pm$5.1$\pm$3.5 MeV, and
$\Gamma_{\gamma\gamma}(\eta_{c})$=7.2$\pm$0.8$\pm$0.7$\pm$2.2(br) keV (the last error is due to the uncertainty in the branching ratio 
${\cal B}(\eta_{c} \to  K_{S}^{0}K^{\pm}\pi^{\mp}$)),
which are in agreement with our previously reported values.
The two-photon width from CLEO III data is
$\Gamma_{\gamma\gamma}(\eta_{c})$=7.5$\pm$0.5$\pm$0.5$\pm$2.3(br) keV. The average of the two results is
$\Gamma_{\gamma\gamma}(\eta_{c})$=7.4$\pm$0.4$\pm$0.5$\pm$2.3(br) keV.

To summarize, in independent analyses of CLEO II and CLEO III
data sets for the reaction
$e^{+}e^{-} \rightarrow e^{+}e^{-}(\gamma\gamma)\rightarrow e^{+}e^{-}
(\eta_{c}^{\prime})\rightarrow e^{+}e^{-} (K_{S}^{0}K^{\pm}\pi^{\mp})$,
we see clear evidence for the excitation of the  $\eta_{c}$(1$^{1}S_{0}$),
and another resonance which we assign to $\eta_{c}^{\prime}$(2$^{1}S_{0}$).
We combine the separate results of CLEO II and CLEO III presented in Table I
to obtain the following as our final results

\begin{center}
$M(\eta_{c})$=
2981.8$\pm$1.3$\pm$1.5 ~ ${\rm MeV,}$ ~~~
\vspace*{0.1cm}

$\Gamma(\eta_{c})$=24.8$\pm$3.4$\pm$3.5~ ${\rm MeV},$~~~~~~~~~~ 
\vspace*{0.1cm}

$M(\eta_{c}^{\prime})$=
3642.9$\pm$3.1$\pm$1.5 ~ ${\rm MeV,}$ ~~~
\vspace*{0.1cm}

$\Gamma(\eta_{c}^{\prime})$=6.3$\pm$12.4$\pm$4.0
${\rm ~MeV, ~~~or}$~~~~

~~~~~~~~~~~~~~~~$\le$ 31${\rm ~MeV~(90\%~CL),}$~~~~~~~~~~~~~~~~~~~
\vspace*{0.1cm}

$R(\eta_{c}^{\prime}/\eta_{c})$=0.18$\pm$0.05$\pm$0.02.~~~~~~~~~~~~~  
\end{center}

Using the known masses of the $J/\psi$ and $\psi^{\prime}$~\cite{pdg2002}, 
and combining statistical and systematic errors in quadrature, 
these correspond to $\Delta M_{hf}(1S)=115.1 \pm 2.0$ MeV, and 
$\Delta M_{hf}(2S)=43.1 \pm 3.4$ MeV. 

Assuming that the branching fractions for 
$\eta_{c}$ and $\eta_{c}^{\prime}$ decays to  $K_{S}K\pi$
are equal~\cite{barnes}, and using the average value of 
 $\Gamma_{\gamma\gamma}(\eta_{c})$ as  obtained above,
our result for $R$ leads to the first estimation of
$\Gamma_{\gamma\gamma}(\eta_{c}^{\prime})$=1.3$\pm$0.6 keV.

As mentioned earlier, all new measurements contradict
the earlier Crystal Ball identification of  $\eta_{c}^{\prime}$
with a mass of 3594$\pm$5 MeV,  and therefore
$\Delta M_{hf}(2S)=92\pm5$ MeV. 
The present results reduce this hyperfine splitting by nearly a factor
two. We hope that this will lead to a reexamination of the 
$c\bar{c}$  hyperfine interaction in the confinement region,
as well as coupled-channel effects for $^{3}S_{1}$ and $^{1}S_{0}$ states.
 
We gratefully acknowledge the effort of the CESR staff
in providing us with
excellent luminosity and running conditions.
This work was supported by
the National Science Foundation,
the U.S. Department of Energy,
the Research Corporation,
and the
Texas Advanced Research Program.


\begin{thebibliography}{99}
\bibitem{errors} Whenever two or more uncertainties are listed,
the first is statistical and the second is systematic.
\end{thebibliography}
\end{document}